\documentclass[10pt, sigconf, a4paper]{IEEEtran}

\def\BibTeX{{\rm B\kern-.05em{\sc i\kern-.025em b}\kern-.08emT\kern-.1667em\lower.7ex\hbox{E}\kern-.125emX}}

\usepackage{algorithmic}
\usepackage{algorithm}

\usepackage{url}
\usepackage{multirow}

\usepackage[justification=justified,font=small,labelfont=bf]{caption}
\usepackage[labelfont=normalfont]{subcaption}
\usepackage{epsfig}
\usepackage{graphicx}
\usepackage{amsmath}
\usepackage{amssymb}
\usepackage{color}
\usepackage{setspace}
\usepackage{cite}
\usepackage{hhline}   
\usepackage{enumerate}
\usepackage{flushend}
\usepackage{graphics}
\usepackage{tikz}
\usepackage{array}
\usepackage{animate}
\usepackage{titlesec}

\definecolor{darkgreen}{RGB}{11,140,21}

\renewcommand{\baselinestretch}{1}
\titlespacing*{\section}{0pt}{4pt}{4pt}
\titlespacing*{\subsection}{0pt}{4pt}{2pt}
\titlespacing*{\subsubsection}{0pt}{4pt}{2pt}
\begin{document}

\title{MAVIREC: ML-Aided Vectored IR-Drop \\Estimation and Classification}
\author{Vidya A. Chhabria$^1$, Yanqing Zhang$^2$, Haoxing Ren$^2$, Ben Keller$^2$, Brucek Khailany$^2$, and Sachin S. Sapatnekar$^1$\\ $^1$University of Minnesota; $^2$NVIDIA Corporation}


\maketitle

\begin{abstract}  
Vectored IR drop analysis is a critical step in chip signoff that checks the power integrity of an on-chip power delivery network. Due to the prohibitive runtimes of dynamic IR drop analysis, the large  number of test patterns must be whittled down to a small subset of worst-case IR vectors. Unlike the traditional slow heuristic method that selects a few vectors with incomplete coverage, MAVIREC leverages machine learning techniques---3D convolutions and regression-like layers---for fast analysis, recommending a larger subset of test patterns that exercise worst-case scenarios. In under 30 minutes, MAVIREC profiles 100K-cycle vectors and provides better coverage and accuracy recommendations than a state of the art industrial flow. Further, MAVIREC's IR drop predictor shows 10X speedup with under 4mV RMSE relative to an industrial flow.
\end{abstract}

\bstctlcite{IEEEexample:BSTcontrol}
\section{Introduction}
\label{sec:intro}
\begin{figure}[tb]
\centering
\includegraphics[width=9cm]{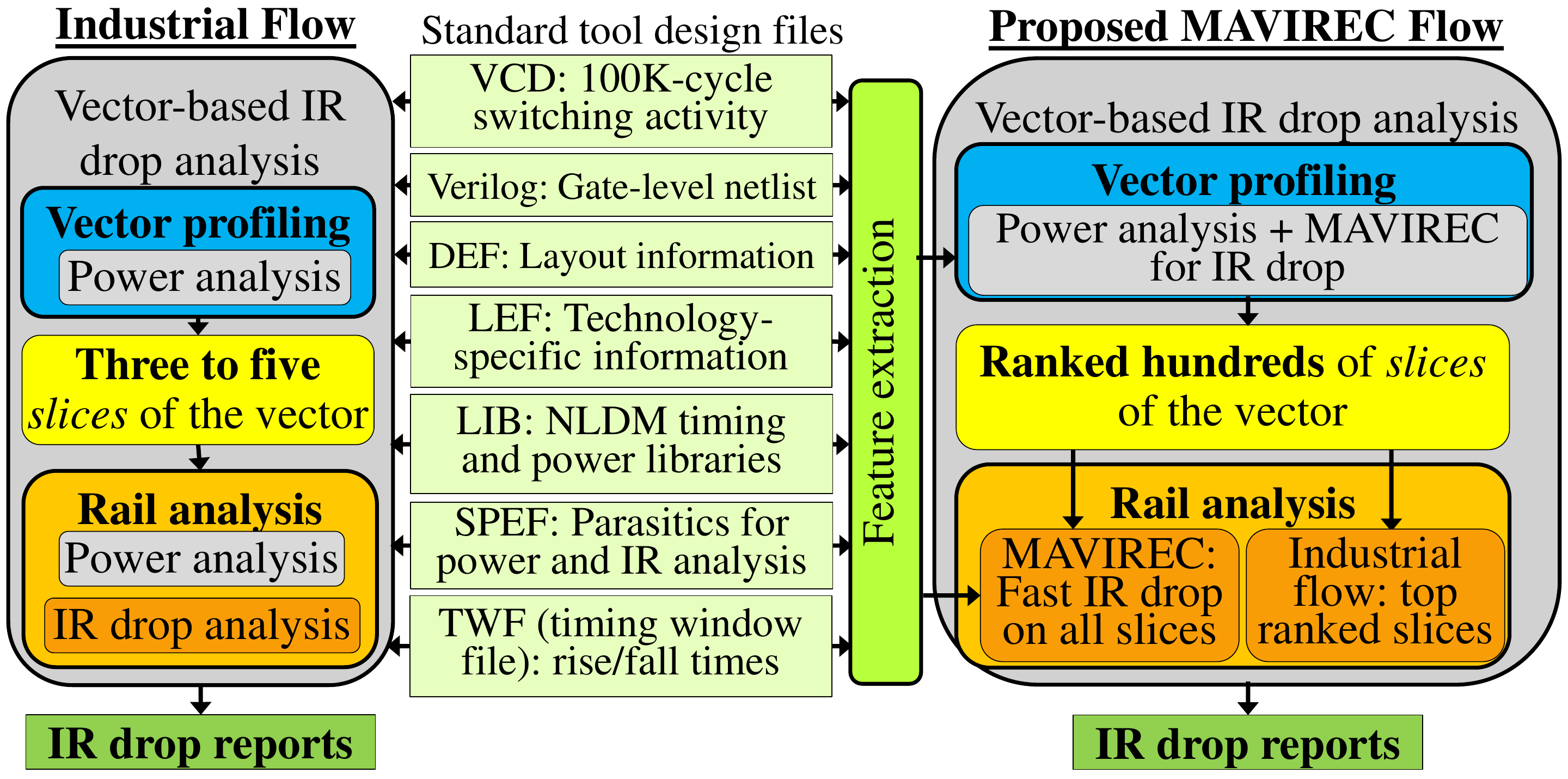}
\caption{An industrial flow (left) vs. the MAVIREC flow (right).}
\label{fig:flow}
\end{figure}

\begin{figure}[tb]
 \vspace{-1.0em}
\centering
\includegraphics[width=5.75cm]{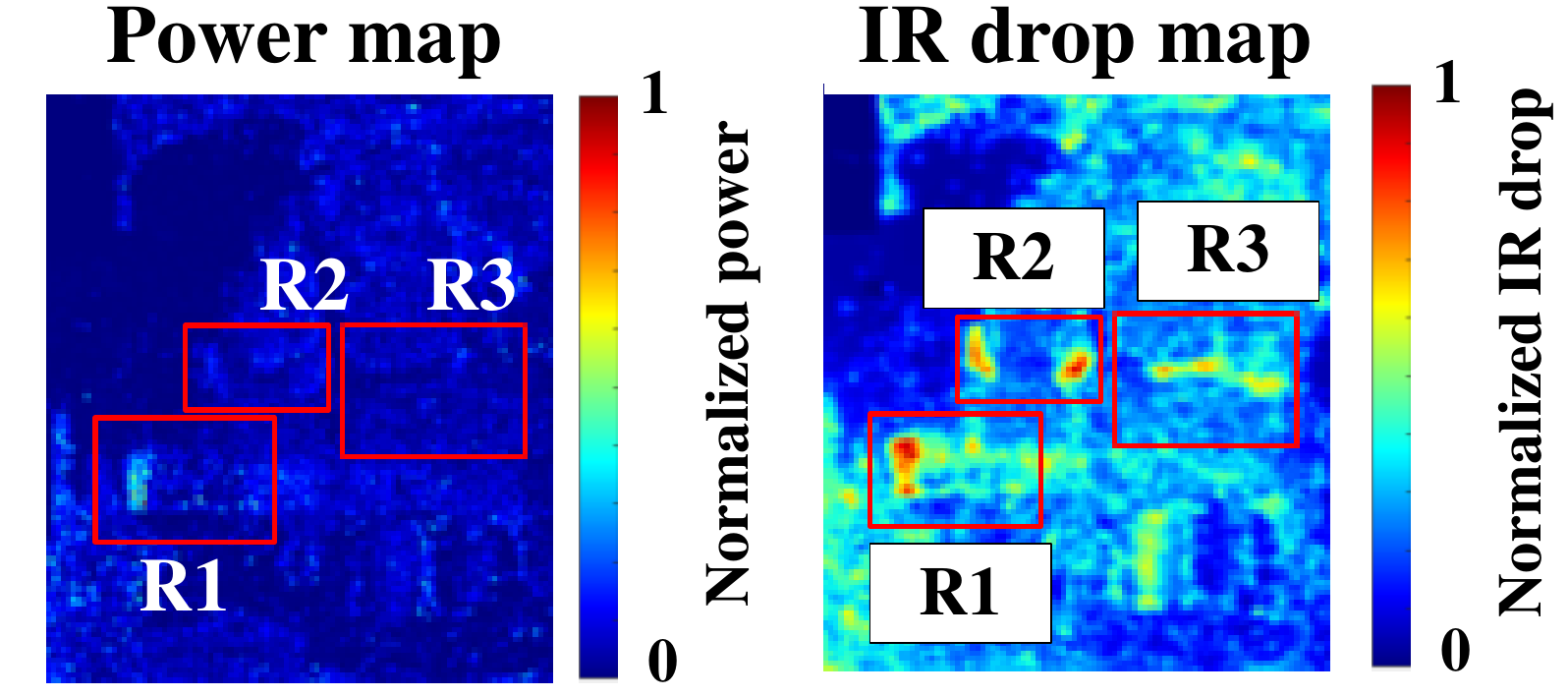}
\caption{An average power map and its IR drop map for a portion of an industry design. Regions with vastly different average power (R1 and R2) can have similar IR drop, and vice versa (R2 and R3).}
\label{fig:power-ir}
\vspace{-2mm}
\end{figure}

\noindent
IR drop analysis estimates the deviation from the ideal supply voltage to meet IR drop constraints across all corners and scenarios. Today, estimating voltage at each node amounts to solving systems of linear equations with billions of variables, with runtimes of several hours on industrial-scale designs even for static analysis. To model switching patterns under real workloads, designers perform dynamic IR drop simulation for several vectors, each spanning a few hundred thousand cycles. Therefore, IR drop analysis in industry is performed for a subset of switching patterns. This subset contains a small number of worst-case \textit{slices} of all vectors, chosen by a vector profiling flow, where a slice corresponds to continuous multicycle time windows of a vector. Fig.~\ref{fig:flow}~(left) shows the industrial flow for vectored IR drop estimation. The blue box performs vector profiling by scanning the entire 100k-cycle vector and recommends a small number of short (up to 20 cycles) worst-case-average-power slices (yellow box). The slices are then sent to the rail analysis engine (orange) to estimate IR drop for each slice. However, such vector profilers are (1)~\textit{Slow}: extracting the top three slices of a 3.2 million instance design for a vector of 5000 slices takes two hours; (2)~\textit{Approximate}: the heuristics inaccurately sort slices in order of their average power~\cite{avg-power-heuristic-1}: as shown in Fig.~\ref{fig:power-ir}, R1 has higher power than R2, both have similar IR drop. (3)~\textit{Computationally limited}: the number of recommended slices is limited by the computational bottleneck of rail analysis, and often do not cover all IR-critical regions. Thus, there is a need for a fast and accurate IR drop estimator to profile large sets of vectors and accurately recommend worst-case slices of the full vector. 
 
Prior machine learning (ML)-based approaches have addressed PDN synthesis~\cite{openpdn} and analysis problems~\cite{xgboost, incpird, powernet, dac-designer-track, vchhabria21}, but none are suited for vectored dynamic IR drop analysis. \cite{xgboost,incpird} target ECO problems and are not suitable for full chip analysis. While dynamic vectorless full-chip IR drop analysis is addressed in~\cite{powernet}, its inference is slow -- our analysis estimates weeks of runtime for full 100k-cycle vector profiling using this method. A U-Net based solution in~\cite{vchhabria21} provides faster ML inference than~\cite{powernet} but is targeted at static IR drop. Its LSTM-based solution for coarse-grained dynamic \textit{thermal} (not IR drop) analysis will face speed and memory problems for fine-grained vectored IR analysis. Moreover, prior works deliver limited accuracy: e.g.,~\cite{powernet,vchhabria21} provide coarse-grained tile-level IR drops, rather than fine-grained instance-level analysis, and the single IR value per tile limits transferability (or reuse) of the model across designs (see Section ~\ref{sec:mavirec-arch}).

We propose MAVIREC (Fig.~\ref{fig:flow}~(right)): a fast, accurate, and novel vector profiling system that recommends worst-case IR drop switching patterns with high coverage of IR hot spots, using a fast and accurate under-the-hood IR estimator. The estimator uses a novel ML architecture based on U-Nets~\cite{unet} with key changes: (1)~It introduces 3D convolutional layers in the model that capture the sparse temporal switching activities, and (2)~It employs a regression-like output layer that enables IR-drop prediction at instance-level granularity. In addition, the use of direct instance-level features as a part of this regression equation enhances transferability and provides model interpretability (see Section~\ref{sec:mavirec-arch}). MAVIREC features:
\begin{itemize}
    \item A novel tailored U-Net-based ML model for dynamic IR drop estimation, achieving 10$\times$ speedup over industrial flows, at the per-instance (i.e., per-logic-gate) granularity, using 3D convolutions for capturing switching activity. 
    \item A novel recommender system that uses fast ML inference to select hundreds of slices when compared to conventional flow that selects 3--5.
    \item An accurate system for profiling long vectors ($\approx$ 100k-cycles), maximizing regional coverage, on industrial scale designs in 30 minutes (4$\times$ speedup vs. industrial flow).
\end{itemize}

\section{MAVIREC for vectored dynamic IR estimation}
\label{sec:unet}
\noindent
The core engine in the proposed flow in Fig.~\ref{fig:flow} (right), for both the vector profiling and rail analysis, is MAVIREC's ML inference scheme. The scheme first performs feature extraction and then uses a trained model for IR drop estimation.

\subsection{Feature extraction}
\label{sec:features}
\begin{figure}[tb]
\centering
\includegraphics[width=9cm]{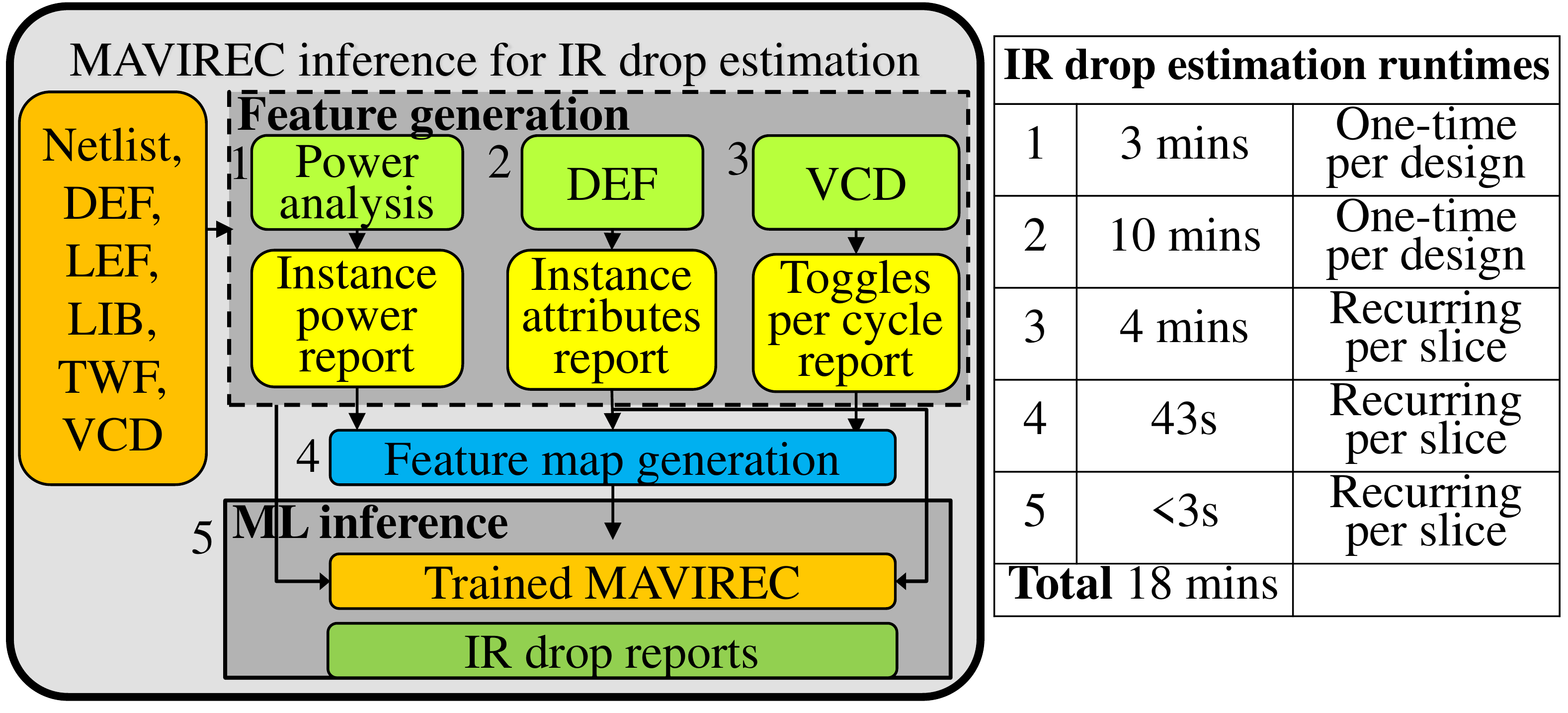}
\caption{MAVIREC ML inference for fast IR drop analysis.}
\label{fig:inference-flow}
\end{figure}

\noindent
The MAVIREC ML model uses two types of features, differentiated by the method in which they are represented:
\begin{itemize}
    \item Instance-level power and effective distance to via stack
    \item Spatial and temporal full-chip tile-based power maps, and spatial time-invariant resistance maps
\end{itemize}
Table~\ref{tbl:features} lists the features used in our ML model, encapsulating power information, the local PDN topology, and switching activity. We first extract these features from the design environment.  Fig.~\ref{fig:inference-flow} shows our two-part process of extracting instance-level features (dotted black box) and generating 2D spatial and temporal feature maps (blue box).

\noindent
{\bf Instance-level feature generation} \label{sec:feat_gen} We run three standard tools highlighted in the green and yellow boxes of Fig.~\ref{fig:inference-flow} to extract instance-level information from the design:
\begin{enumerate}
    \item a power analysis tool to generate a per-instance switching ($p_s$), internal ($p_i$), and leakage power ($p_l$) report.
    \item a parser tool to extract from the DEF (a)~instance locations (b)~effective distance of an instance to power rail via stacks in the immediate (5$\mu$m) neighborhood. The effective distance metric is: $r^{-1} = d_1^{-1} + ...+d_V^{-1}$ where $d_i$ is distance from the instance of the $i^{th}$ of $V$ via stacks. 
    \item a tool that extracts the times when instances toggle, from the VCD file (industry-standard format) for each slice. 
\end{enumerate}
\noindent

\noindent
 Similar to~\cite{powernet}, in addition to $p_l$, $p_s$, $p_i$, and $r$, there are three other instance-level features which are derived from these. The toggle rate scaled and total power are given by $p_r = p_l + \tau_i(p_s + p_i)$ and $p_{tot} = p_l + p_s + p_i$ respectively, where $\tau_i$ is the average toggle rate of the instance in the specified slice. The overlap power $p_{ol}$ sums up the $p_r$ of all neighboring instances that share the same timing window.

\begin{table}
\centering
\caption{MAVIREC's features for dynamic IR prediction where the lowercase symbols are instance-level and uppercase are at the tile-level.}
\label{tbl:features}
\resizebox{\linewidth}{!}{%
\begin{tabular}{||l|l||} 
\hhline{|t:==:t|}
\multicolumn{2}{||c||}{\textbf{List of all features}} \\ 
\hhline{|:==:|}
Internal power: $p_i$, $P_i$ & Overlap power: $p_{ol}$, $P_{ol}$ \\ 
\hline
Leakage power: $p_l$, $P_l$ & Total power: $p_{tot}$, $P_{tot}$ \\ 
\hline
Switching power: $p_s$, $P_s$ & Effective distance: $r$, $R$ \\ 
\hline
Toggle rate scaled power: $p_r$, $P_r$ & Toggle power at each time step: $p_t$, $P_t$ \\ 
\hhline{|:==:|}
\multicolumn{2}{||c||}{\textbf{Ground truth training labels}} \\ 
\hhline{|:==:|}
\multicolumn{2}{||l||}{Instance-level IR
  drop: $IR_i$, NA (heat map only)} \\
\hhline{|b:==:b|}
\end{tabular}
}
\end{table}

\noindent
{\bf Generation of 2D spatial and temporal maps as features} \label{sec:mapgen} The map creation step performs a spatial and temporal decomposition (Fig.~\ref{fig:inference-flow} (blue box)) of the per-instance features in a similar manner as~\cite{powernet}.
For the spatial-tile based features, we associate the location of each instance with its corresponding power/effective distance attributes to create 2D distributions of these attributes at a fixed granularity (\textit{tile} size). We choose a tile size of $W_t = 2.5 \mu$m $\times$ $L_t = 2.5\mu$m; a design of size $W_c$ $\times$ $L_c$ corresponds to an ``image'' of $ W (= W_c/w_t) \times L (= L_c/l_t)$ tiles.  For each instance-level feature in Table~\ref{tbl:features}, its tile-based counterpart adds per instance power-related features, and takes the maximum effective distance over all instances, in the tile.

For dynamic IR drop analysis, we also consider temporal power maps. To generate $p_t$ (the power of an instance at each time step) we divide the $n$-cycle time window into $n \times t$ time steps, where $n$ is usually predetermined by the designer as the window size of interest (we use $n=20$) and $t$ is a hyperparamter to be tuned. It was observed that $t=5$ provided the best results with MAVIREC. The power of an instance at time step $j$ is given by: $p_t(j) =\sum_{j = 1}^{100}p_l + b_j(p_i + p_s)$
\noindent
where the Boolean variable $b_j$ is 1 only if the instance toggles at time step $j$. Thus, we create time-decomposed power maps at each time step using the toggle information and timing windows. 

All features in Table~\ref{tbl:features} are normalized between 0 and 1 and are inputs to the trained ML model. Normalization scales the input by a predetermined constant that is defined for the technology, e.g., the maximum supply voltage or current.

\begin{figure}[tb]
\centering
\includegraphics[width=9cm]{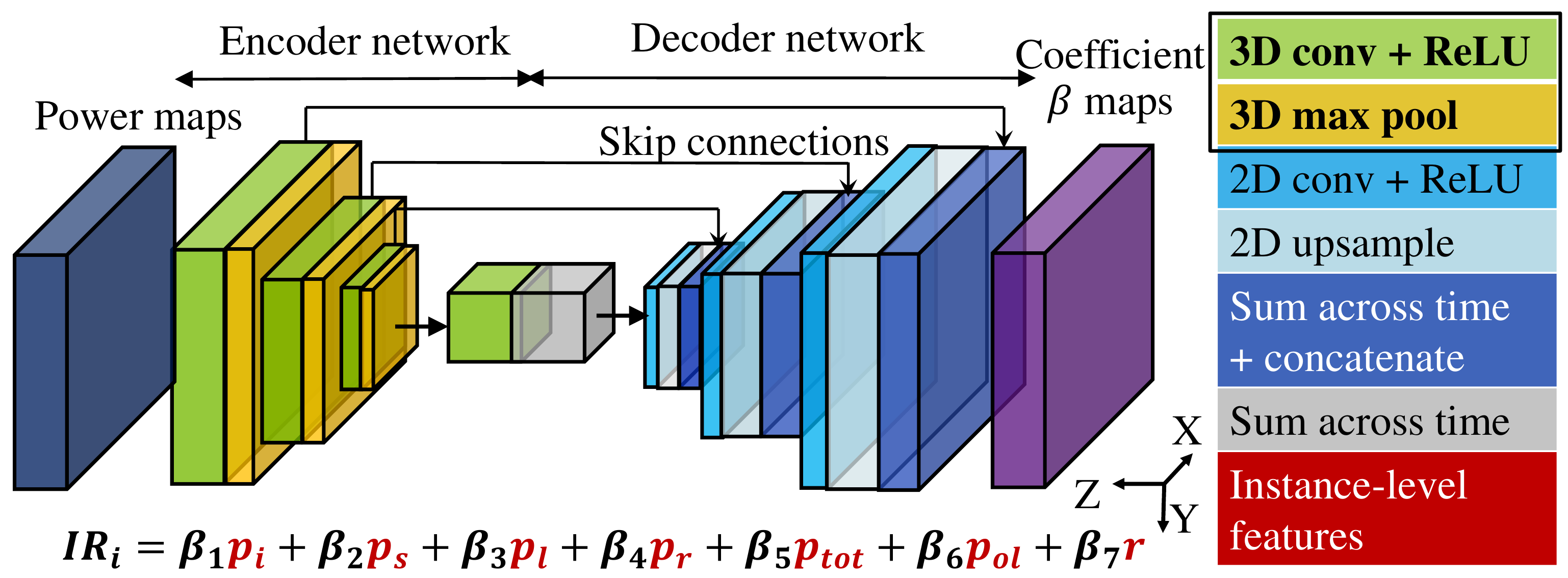}
\caption{MAVIREC: 3D convolution and regression model based U-Net architecture for dynamic IR drop prediction.}
\label{fig:mavirec-arch}
\end{figure}

\subsection{MAVIREC architecture}
\label{sec:mavirec-arch}

\noindent
Fig.~\ref{fig:mavirec-arch} shows the structure of MAVIREC for vectored dynamic IR drop estimation, with layer descriptions provided in the legend. It consists of two subnetworks for (i) encoding (downsampling) and (ii) decoding (upsampling), with skip connections between the networks. The skip connections use a concatenation layer to incorporate information from an early layer into a deeper stage of the network, skipping intermediate layers, and appends it to the embedding along the z-direction.  This architecture is based on fully convolutional (no fully connected layers) U-Net models~\cite{unet} used for image segmentation.  The fully convolutional nature of U-Nets makes them fast and they are designed to be input-image-size-independent. The convolutional and max-pool layers of the U-Net capture local and global spatial neighborhood features.

The work in~\cite{vchhabria21} uses U-Nets for estimating {\em static} IR drop. However, for reasons listed in Section~\ref{sec:intro}, this model is not suitable for vectored dynamic IR drop analysis. In addition, typical for real workload vectors, cells switch at only a few spatial and temporal locations. This makes the temporal power map $P_t$ extremely sparse and difficult to capture in the zero-dominant data. As shown in Section~\ref{sec:results-ir}, merely accounting for these sparse temporal features is not sufficient, and the ML architecture must be able to capture these sparse local changes accurately. There are two key differences from the U-Net in~\cite{vchhabria21} that are crucial for overcoming its limitations:

\begin{itemize}
\item {\em 3D convolutional layers} (green layers) in the encoding path captures temporal simultaneous switching activity.
\item {\em regression-like layer} at the end of the decoder that incorporates instance-level input features and the instance-level IR drop {\bf \textit {IR}$_i$} (equation in Fig.~\ref{fig:mavirec-arch}) using $\boldsymbol{\beta_i}$, the coefficient matrix predicted by the U-Net like structure.
\end{itemize}
{\bf 3D convolutional layer in the encoder}\label{sec:3dconv_enc} Unlike a 2D convolutional layer that uses all input channels during convolution, a 3D convolutional layer restricts the number of channels to the specified filter size in the channel dimension, thus considering only a small local window of channels. When all the temporal power maps are taken together as channels in a regular 2D convolutional layer, due to zero-dominance in the data, the model fails to capture key non-zero toggle activity regions and time steps. Intuitively, a small local window of channels which a 3D convolutional layer considers would better capture simultaneous and sparsely-distributed switching activity. Therefore, MAVIREC uses a $3\times3\times3$ filtered 3D convolutional layer in the encoding path instead of a regular $3\times3$ 2D convolutional layer as in U-Net (green layer in Fig.~\ref{fig:mavirec-arch}). MAVIREC has $n \times t + 7$ tile-based channels (Table~\ref{tbl:features}):
\begin{itemize}
\item $n\times t$ temporal power maps ($P_t$) to the encoder network
\item 7 tile-based spatial features ($P_i$, $P_l$, $P_s$, $P_{r}$, $P_{ol}$, $P_{tot}$, $R$)
\end{itemize}
where $n\times t$ represents the number of time steps (Sec.~\ref{sec:features}).

We will show in Section~\ref{sec:results-ir} that a 2D convolution model does not accurately capture simultaneous switching compared to 3D convolution.  MAVIREC consists of four 3D convolutional layers and three 3D max pool layers in the encoder network and four 2D convolutional layers and three upsampling layers in the decoder network. Since the decoding path uses 2D convolutions, the interface between the 3D embedding in the encoder and the 2D embedding in the decoder sums up the embedding along the temporal dimension (dark blue boxes in Fig.~\ref{fig:mavirec-arch}) through concatenation/skip connections. 

\noindent
{\bf Regression-like layer in the decoder}\label{sec:reg-layer}
To enable IR drop prediction at a per-instance level, MAVIREC leverages a  regression-like layer at the end of the decoder path that uses instance-level input features and multiplies them with the predicted coefficients ($\beta_i$) by the U-Net-like structure in Fig.~\ref{fig:mavirec-arch}. The predicted coefficients are based on the $n \times t + 7$ spatial and temporal tile-based channels as input.
The coefficient predicted for every tile is then multiplied with the per-instance feature values defined in Sec.~\ref{sec:features}. This architecture provides three key advantages over prior ML-based solutions~\cite{powernet,vchhabria21}:
\begin{enumerate}
    \item {\em Improved transferability} compared to prior art, as the model uses both instance-level features directly and tile-level features to predict IR drop at a per-instance level. 
    The instance-based features help capture fine-grained variations in the data that is otherwise lost due to the averaging nature of U-Net convolutions. Instead of learning the IR drop values directly as in~\cite{powernet},~\cite{vchhabria21}, the U-Net-like structure learns the relationship ($\boldsymbol{\beta_i}$) between the features and the IR drop values, which is universal across designs.
    \item {\em Improved instance-level IR drop predictability} compared to prior art, which is useful for applications such as IR-aware STA and instance-based IR drop mitigation.
    \item {\em Model interpretability} as the predicted coefficients are the weights associated with each feature. The coefficients correspond to feature sensitivity, and allow designers to assess the root cause of an IR drop violation. 
\end{enumerate}
A trained MAVIREC model is reusable, without the need for retraining when faced with new designs/vectors for a given technology.
Although the prediction is on an instance level granularity, it is not necessary to loop through all instances to estimate the IR drop. Instead, we define a location matrix that creates a mapping between each instance and its corresponding tile-based coefficient. This mapping eliminates the loop by using element-wise matrix multiplication which is accelerated on GPUs (details have been omitted due to space constraints). 
\section{MAVIREC for vector profiling}
\label{sec:vector-profiling}

The input vectors in real workloads can typically have hundred of thousands of cycles, which corresponds to $\sim$5000 slices. These slices of 20 cycle windows each form inputs to the rail analysis step for IR drop estimation. Given the runtimes of industrial flow rail analysis, it is near-impossible to run rail analysis on all slices.  Using a brute force approach, where MAVIREC is used as a fast IR drop estimator, is not a viable solution either as each inference involves extracting temporal features at recurring runtime costs (see Fig.~\ref{fig:inference-flow}) for each of the 5000 slices. This calls for techniques to select a large set (70--200) of candidate slices that represent the design and worst-case IR drop vector. In contrast, industrial flows are limited to 3--5 slices due to IR analysis run-times.

\begin{figure}[tb]
\centering
\includegraphics[width=9cm]{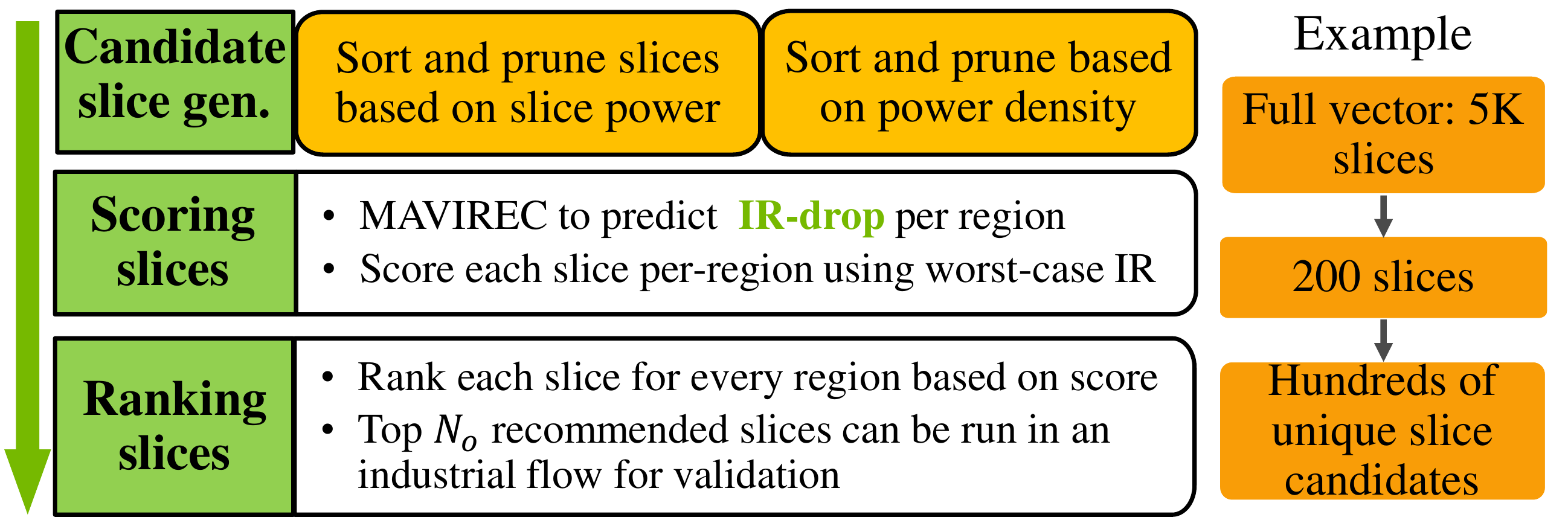}
\caption{MAVIREC vector profiling which uses ML inference under the hood to generate hundreds of ranked candidates.}
\label{fig:vector-profiling-flow}
\end{figure}

Fig.~\ref{fig:vector-profiling-flow} and Algorithm~\ref{algo:vector-profiling}
detail MAVIREC's vector profiling flow. The inputs are per-instance leakage, switching and internal power $p_l$, $p_s$, and $p_i$, region size $w \times l$, chip size $W_c \times L_c$, the number of slices to select at each of the two stages of candidate generation given by $N_a$ and $N_r$, and $N_o$ is the number of top rank slices to report at the end of the profiling. 
 The number of regions in the design is given by $W_r \times L_r$ where $W_r = W_c/w$ and $L_r = L_c/l$.
\begin{algorithm}[!h]
    {\scriptsize
    {\bf Input}: Per instance internal, switching, and leakage powers $p_i$, $p_s$, $p_l$; Per instance toggle count $T_c$; Number of slices to select $N_a$, $N_r$, and $N_o$; Region size $w \times l$; Chip dimensions $W_c \times L_c$; List of all candidate slices in vector $C_N$; \\
    {\bf Output}: Candidate slices for worst-case IR drop $C_{N_o}$\\
    \hspace*{8.5mm} IR drop map $IR_{chip, N_o}$ of the $C_{N_o}$ candidates
    \begin{algorithmic}[1]
        \STATE $W_r = \frac{W_c}{w}$ and $L_r= \frac{L_c}{l}$ \\
	    \STATE \textit{// Candidate slice generation starts here} \\
	    \label{algo:can-gen-start}
	    \STATE \textit{// Stage 1: sort and eliminate based on average slice power} \\
	    \FOR {each slice $c$ in $C_N$}\label{algo:stage1-start}
	         \STATE {$P_{slice}[c]= \sum_{\forall \text {instances}}^{} \left [ p_l + \frac{T_c[c]}{20} \left ( p_s + p_i \right ) \right ]$}
	         \label{algo:avg-slice-power}
	   \ENDFOR
	   \STATE {$C_{N_a}$ = Top $N_a$ candidate slices with highest value in $P_{slice}$}
	   \label{algo:stage1-end}
	   \STATE \textit{// Stage 2: sort and eliminate based on average regional power} \\
	   \label{algo:can-slice-power}
	   \FOR {each region $r_i$ in $W_r \times L_r$}
	   \label{algo:stage2-start}
	        \FOR {each candidate slice $c$ in $C_{N_a}$}
	            \STATE {$P_{R}[r_i][c]=\sum_{\forall \text {instances} \in r_i}^{} \left [ p_l + \frac{T_c[c]}{20} \left ( p_s + p_i \right ) \right ]$ }
	            \label{algo:avg-region-power}
	         \ENDFOR
	       \STATE {$C_{N_r}[r_i]$ = Top $N_r$ candidate slices with highest value in $P_R[r_i]$}
	       \label{algo:can-region-power}
	   \ENDFOR
	   \STATE {$C_{N_c}$ = Unique candidates from $C_{N_r}$}
	    \label{algo:stage2-end}
	   \label{algo:can-gen-end}
	   \STATE \textit {// Candidate scoring and ranking starts here}
	   \FOR {each slice $c$ in $C_{N_c}$}  \label{algo:score-rank-start}
	        \STATE {$F[c]$ = Feature\_extraction($c$)}
	        \label{algo:MAVIREC-features}
	        \STATE {{\textit {IR}$_{chip}$} = ML\_inference($F[c]$)} \\
	        \label{algo:inference}
	        \FOR{each region $r_i$ in $W_r \times L_r$}
	            \STATE {\textit {IR}$_{score}[r_i][c]$ = max(\textit{IR}$_{chip}[c]$) in $r_i$}
	            \label{algo:max-region-ir}
	        \ENDFOR
	   \ENDFOR
	   \STATE $n = 1$
	   \WHILE{$n \leq N_o$}
	            \STATE $r_i$, $C_{N_o}[n]$ = Top $N_o$ unique candidates, regions with highest \textit{IR}$_{score}$ values\label{algo:rank-unique-reg-slice}
	            \STATE {\textit {IR}$_{chip,N_o}[n] =$ \textit{IR}$_{score}[x][C_{N_o}[n]]$  $\forall$ $x$ $\epsilon$  $W_r \times L_r$}
	            \label{algo:ir-final}
	            \STATE{$n$++}
	            \label{algo:ir-rank}
	  \ENDWHILE
	  \label{algo:score-rank-end}
    \end{algorithmic}
    \caption{MAVIREC's algorithm for vector profiling}
    \label{algo:vector-profiling}
    }
\end{algorithm}

Inspired by classical ML recommender system~\cite{youtube} nomenclature, our method (Fig.~\ref{fig:vector-profiling-flow}) is a three-step process which consists of 
(i) candidate generation (two stages), (ii) scoring, and (iii) ranking. At each stage of candidate generation, we prune out slices based on average power first and regional-power next. We score the remaining slices using the worst-case IR drop as a metric and rank them using a technique that maximizes regional coverage.  
Fig.~\ref{fig:vector-profiling-flow} displays an example of candidate generation where a number of slices are pruned out of the set at each stage.

\noindent
{\bf Candidate slice generation:} This step of the vector profiling flow consists of two stages (Lines~\ref{algo:stage1-start}--\ref{algo:can-gen-end}).  

\noindent
\underline{First stage} In each of our vectors, we observe that there are thousands of slices with near-zero switching activity. Therefore, in the first stage of candidate generation we sort and prune the set of slices based on the average power of each slice, as shown in lines~\ref{algo:stage1-start}--\ref{algo:stage1-end}. Pruning at this stage saves time by preventing feature extraction for slices that are obviously not IR-critical. The pruned and sorted list, $C_{N_a}$, has hundreds of potential candidates to move on to the next stage of our flow (much more than 3--5 in industrial flows).

\noindent
\underline{Second stage} Since the IR drop of an instance depends on the switching activity of an instance and its neighborhood, power per region is vital for IR drop. Therefore, in the second stage of candidate generation (lines~\ref{algo:stage2-start}--\ref{algo:stage2-end}), we calculate the power per region in the design and then rank the $C_{N_a}$ candidate slices in each region. We then extract the top $N_r$ from each region. This results in a list, $C_{N_r}$, of $N_r \times W_r \times L_r$ candidate slices and $C_{N_c}$ is the list of $N_c$ unique candidates from $C_{N_r}$. 

\begin{figure}[bt]
\centering
\includegraphics[width=8.2cm]{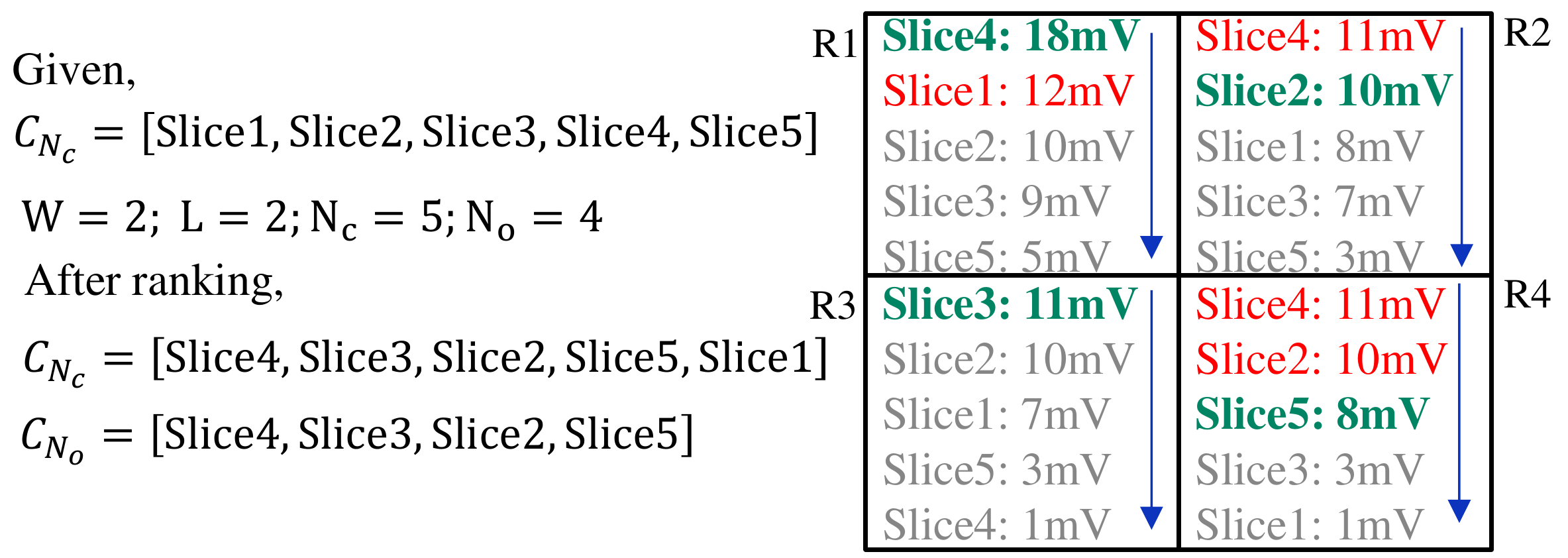}
\caption{Scoring and ranking scheme for $C_{N_c}$ candidates.}
\label{fig:score-rank-ex}
\end{figure}

\noindent
{\bf Candidate slice scoring and ranking} At this step, we score each of the $N_c$ (hundreds) generated candidates based on worst-case IR drop of each slice in each region. For this, we take each candidate slice $C_{N_c}$ and generate all the required features required  (Table~\ref{tbl:features}) for the ML inference in line~\ref{algo:MAVIREC-features}. The inference engine in MAVIREC is then used to generate a full-chip IR drop as detailed in line~\ref{algo:inference} and Section~\ref{sec:mavirec-arch}. The worst-case IR drop across the chip is used to score each slice. This results in $N_c \times W_r \times L_r$ score values in \textit{IR}$_{score}$(line~\ref{algo:max-region-ir}). We then rank each candidate based on the worst-case IR drop and record its corresponding region (line~\ref{algo:rank-unique-reg-slice}). Only those unique $N_o$ candidates that correspond to the worst-case IR drop of a previously uncovered region are stored in $C_{N_o}$.

Fig.~\ref{fig:score-rank-ex} shows the scoring and ranking scheme for a demonstrative example of a chip with four regions R1--R4 and $N_c=5$ generated candidate slices. For each region, the slices are ranked in decreasing order of the score/worst-case IR drop in the region. The unique slice from each region with the highest score is selected. The worst-case IR drop of the design is 18mV and corresponds to Slice 4 in R1. The next highest IR drop is Slice 1, with 12mV in R1. However, since R1 has already been covered, we do not need to report Slice 1. The next highest IR drop is Slice 3 in R3 and Slice 4 in R2 with 11mV. In this case, Slice 4 has already been covered by R1, which makes Slice 3 the next highest rank. We repeat the same process until we cover all $W_r \times L_r$ regions. 
In this way, we ensure to report candidates based on worst-case IR drop while maximizing regional coverage. We also report the IR drop maps of $C_{N_o}$ candidate list (line~\ref{algo:ir-final}). 

\noindent
{\bf Computational complexity}
Our candidate generation step is of the order of O($N \log N_a$) + O($W_r L_r N_a \log N_r$), for the first and second stage respectively. The $\log N_a$ and $\log N_r$ come from keeping track of the top $N_a$ and $N_r$ candidate slices which are small user-defined upper-bounded numbers that do not change across designs and tests. We use $N_a = 200$ and $N_r = 5$ which changes the complexity to O($N$) + O($W_rL_r$). In practice, $N$, the number of slices in the vector, are of the order of the hundreds of thousands while $W_rL_r$, the number of regions in the design, is in the order of thousands. Therefore, we have a complexity of O($N$), linear in the number of slices in the vector.

\section{Results and Discussions}
\label{sec:results}
\noindent
In our experiments, we use four industrial designs, D1--D4, implemented in a sub-10nm FinFET technology and three multi-cycle vectors, T1--T3, per design. Table~\ref{tbl:designs-all} summarizes the designs and tests used in our training set. The data available to us were taped out designs in an industrial setting with low IR drop values. Therefore, we use a threshold of 8mV to classify IR-critical regions. 
The model is trained end-to-end using golden per-instance IR drop labels obtained from an industrial flow using commercial tools. We evaluate MAVIREC with \textit{leave one out cross validation}, where in each training run one design and its tests are omitted from the training set.
These are implemented in a PyTorch 1.6~\cite{pytorch} framework on an 8-core CPU machine with 256GB RAM and one NVIDIA Tesla V100GPU with 32GB RAM. We use an ADAM optimizer~\cite{adam} for training and use an L$_2$ regularizer to prevent overfitting. The model takes 9 hours to train and is a one-time cost per technology. The trained model is transferable across both designs and vectors.  
 

\begin{table}
\centering
\caption{Summary of designs and vectors used in MAVIREC experiments with a tile size of $2.5\mu$m $\times$ $2.5\mu$m.}
\label{tbl:designs-all}
\resizebox{\linewidth}{!}{%
\begin{tabular}{||c|c||c|c|c|c|c|c||} 
\hhline{|t:==:t:======:t|}
\multirow{2}{*}{ \textbf{Design} } & \multirow{2}{*}{\begin{tabular}[c]{@{}c@{}}\textbf{\#inst.}\\\textbf{(mill.)} \end{tabular}} & \multicolumn{2}{c|}{\textbf{T1} } & \multicolumn{2}{c|}{\textbf{T2} } & \multicolumn{2}{c||}{\textbf{T3} } \\ 
\cline{3-8}
 &  & \begin{tabular}[c]{@{}c@{}}\textbf{\%IR}\\\textbf{-critical}\\\textbf{regions} \end{tabular} & \begin{tabular}[c]{@{}c@{}}\textbf{Toggle }\\\textbf{rate} \end{tabular} & \begin{tabular}[c]{@{}c@{}}\textbf{\%IR}\\\textbf{-critical}\\\textbf{regions} \end{tabular} & \begin{tabular}[c]{@{}c@{}}\textbf{Toggle }\\\textbf{rate} \end{tabular} & \begin{tabular}[c]{@{}c@{}}\textbf{\%IR}\\\textbf{-critical}\\\textbf{regions} \end{tabular} & \begin{tabular}[c]{@{}c@{}}\textbf{Toggle }\\\textbf{rate} \end{tabular} \\ 
\hline
D1 & 3.26 & 13.66 & 0.054 & 16.04 & 0.062 & 8.98 & 0.041 \\ 
\hline
\textbf{D2}  & 2.19 & 4.60 & 0.043 & 4.24 & 0.040 & 3.59 & 0.038 \\ 
\hline
\textbf{D3}  & 2.18 & 4.28 & 0.040 & 4.31 & 0.038 & 3.88 & 0.038 \\ 
\hline
\textbf{D4}  & 2.43 & 16.88 & 0.085 & 10.68 & 0.087 & 13.00 & 0.089 \\
\hhline{|b:==:b:======:b|}
\end{tabular}
}
\end{table}

\subsection{MAVIREC for IR drop prediction and classification}
\label{sec:results-ir}
{\bf MAVIREC vs. industrial flow for IR drop}
We compare MAVIREC-predicted IR drop against a ground truth IR drop from an industrial flow.
Similar to the industrial flow, MAVIREC predicts IR drop at a per-instance granularity. Table~\ref{tbl:mavirec-redhawk} shows the RMSE and max error for all the designs and tests and it is seen that the RMSE is very small ($<$4mV) and is within reasonable limits for instance-level applications such as cell-level IR drop mitigation and IR-aware STA. 

The rest of the table depicts the performance of MAVIREC as an IR drop hotspot classifier on a per-tile basis, where a tile is considered hot if the average IR drop of all instances in that tile is greater than the threshold (8mV). We consider two different granularities, a $1\times1$ tile ($2.5\mu$m $\times$ $2.5\mu$m) and $6\times6$ tiles ($15\mu$m $\times$ $15\mu$m), and report the accuracy for each design and test. 
We obtain an average accuracy of 93.12\% and 91.22\% at the $1\times1$ and $6\times6$ granularities respectively. At $6\times6$ granularity we have an F1 score of 0.78, which despite the heavily imbalanced minority class ($<$10\% in the dataset), still captures all large hotspots which are of interest to designers. This accuracy outperforms prior art (Section~\ref{sec:results-ir}) and is sufficient for vector profiling (Section~\ref{sec:results-vp}). 

\begin{figure}[tbh]
\centering
\includegraphics[width=\linewidth]{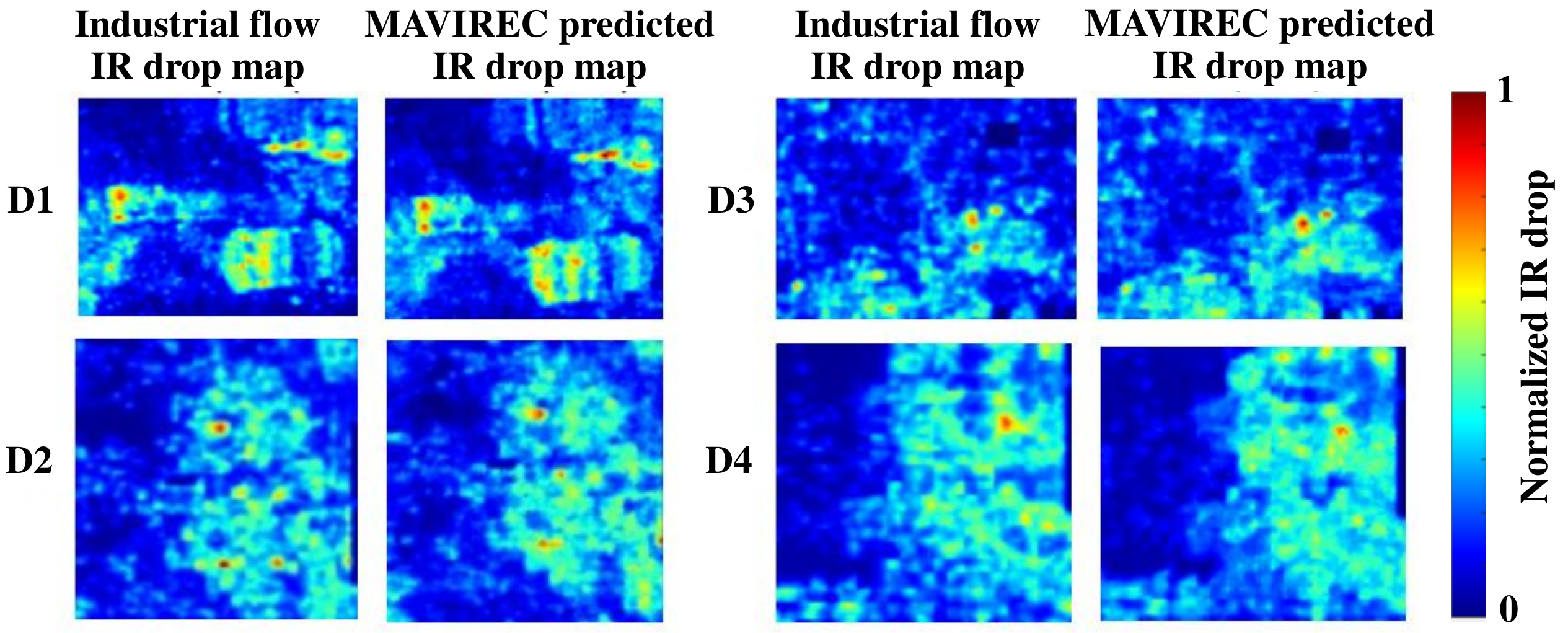}
\caption{Visualization of MAVIREC-predicted IR drop map and industrial flow IR drop map for all four designs.}
\label{fig:maps}
\end{figure}

For a visual comparison of the IR drop hotspot map, we convert the instance-based {\bf IR$_{\bf i}$} to tile-based, taking the mean IR drop of all instances in a ($2.5\mu$m $\times$ $2.5\mu$m) tile. Fig.~\ref{fig:maps} shows the predicted and ground truth IR maps for a section of each design for test T1. The MAVIREC-predicted IR drop map captures all major hotspots relative to the ground truth.

\begin{table}[tb]
\centering
\caption{Performance of MAVIREC ML inference compared to an industrial flow. RMSE and MAE at instance-level granularity and \% accuracy  as a binary classifier at region-based granularity.}
\label{tbl:mavirec-redhawk}
\resizebox{\linewidth}{!}{%
\begin{tabular}{||l|l||r|r||l|l||r|r||} 
\hhline{|t:==:t:==:t:==:t:==:t|}
\multicolumn{1}{||c|}{} & \multicolumn{1}{c||}{} & \multicolumn{1}{c|}{\begin{tabular}[c]{@{}c@{}}\textbf{ RMSE, MAE}\\\textbf{ (mV)} \end{tabular}} & \multicolumn{1}{c||}{\begin{tabular}[c]{@{}c@{}}\textbf{\% Accuracy }\\\textbf{ 1x1, 6x6} \end{tabular}} &  & \multicolumn{1}{c||}{} & \multicolumn{1}{c|}{\begin{tabular}[c]{@{}c@{}}\textbf{ RMSE,MAE}\\\textbf{ (mV)} \end{tabular}} & \multicolumn{1}{c||}{\begin{tabular}[c]{@{}c@{}}\textbf{\% Accuracy }\\\textbf{ 1x1, 6x6} \end{tabular}} \\ 
\hhline{|:==::==::==::==:|}
\multirow{3}{*}{\textbf{D1} } & \textbf{T1}  & 4.44, 33 & 89.09, 84.98 & \textbf{D3}  & \textbf{T1}  & 3.02, 26.8 & 96.26, 93.45 \\ 
\cline{2-8}
 & \textbf{T2}  & 4.61, 25.2 & 85.49, 88.72 &  & \textbf{T2}  & 2.99, 22.4 & 96.12, 92.89 \\ 
\cline{2-8}
 & \textbf{T3}  & 3.91, 30.2 & 92.93, 88.93 &  & \textbf{T3}  & 2.93, 35.1 & 95.96, 92.56 \\ 
\hline
\multirow{3}{*}{\textbf{D2} } & \textbf{T1}  & 3.1, 25.4 & 95.68, 92.73 & \textbf{D4}  & \textbf{T1}  & 3.87, 24.1 & 89.43, 91.07 \\ 
\cline{2-8}
 & \textbf{T2}  & 3.11, 20.6 & 95.86, 92.77 &  & \textbf{T2}  & 3.38, 24.7 & 90.27, 93.62 \\ 
\cline{2-8}
 & \textbf{T3}  & 2.99, 21.8 & 96.08, 92.19 &  & \textbf{T3}  & 3.31, 24.5 & 90.25, 94.55 \\
\hhline{|b:==:b:==:b:==:b:==:b|}
\end{tabular}
}
\end{table}

\noindent 
{\bf MAVIREC versus ML-based IR drop classifiers}
As a comparison to prior art, we implemented a maximum CNN structure similar to PowerNet~\cite{powernet}. To ensure a fair comparison, we make the following changes: (i) As the focus of PowerNet was vectorless IR drop estimation and does not support the use of multi-cycle switching activity, to adapt PowerNet, we take the average toggle rate from the vectors. (ii) PowerNet predicts IR drop on a per-tile ($1\mu$m $\times$ $1\mu$m) basis, while MAVIREC is on a per-instance basis. Therefore, MAVIREC generates a region-based IR drop, like PowerNet, by taking the mean of the predicted-IR drop of all instances in the tile. 

We compare MAVIREC against two baseline ML models: (i) a vanilla U-Net, identical to MAVIREC except that it uses 2D convolutional layers,
and (ii) a max U-Net, identical to PowerNet except that the CNN is replaced with a 2D convolutional U-Net with an output regression layer (Section~\ref{sec:unet}). The vanilla U-Net processes all time steps simultaneously while the max U-Net processes each time step separately. The max U-Net sets the final IR drop to the maximum IR drop for each instance across all time-steps. 
We train all models on the same dataset and tune the hyperparameters for optimal performance.

 \begin{figure}[htb]
\centering
 \includegraphics[width=7.5cm]{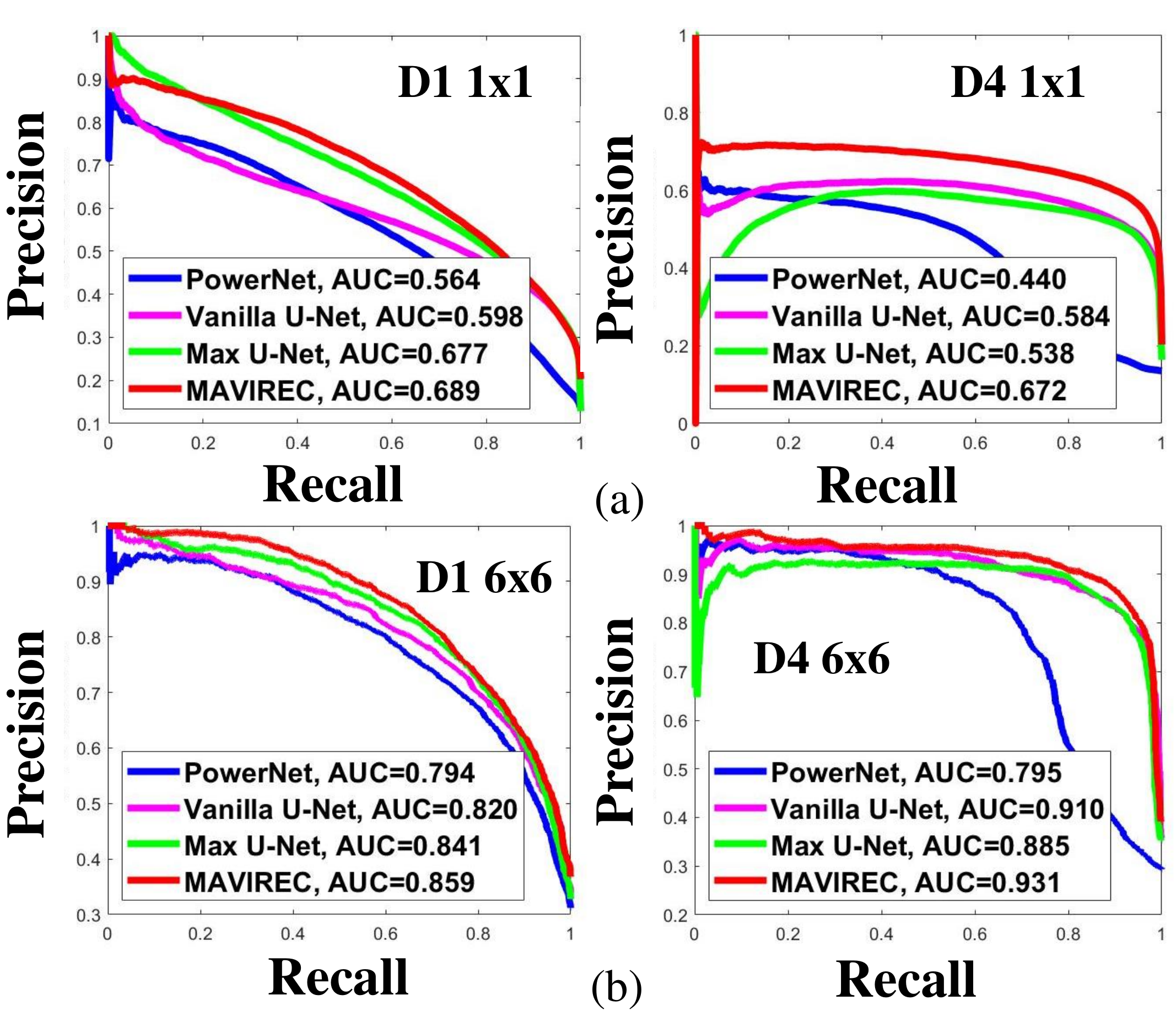}
\caption{Comparison of MAVIREC against PowerNet, Max U-Net, and 2D U-Net. Precision Recall curves showing MAVIREC's ability to predict the minority class on designs D1 and D4 (as representative examples) for a tile size of (a) 1x1 and (b) 6x6 tiles}
\label{fig:pr}
\end{figure}




The area under curve (AUC) of the precision-recall (PR) curve is a single value that demonstrates the ability of the classifier to predict the minority class, in a highly imbalanced dataset. A random classifier has a PR curve equal to the fraction of minority class in the dataset, while for a classifier that always predicts the majority class, the PR curve coincides with the x-axis. 
Fig.~\ref{fig:pr} shows that MAVIREC outperforms both PowerNet and baseline ML models in predicting the IR-critical class across designs with larger AUCs for $1\times1$ (Fig.~\ref{fig:pr}(a)) and $6\times6$ (Fig.~\ref{fig:pr}(b)) tile granularities. The 3D convolutional layers in MAVIREC capture the sparse switching activity of the vectors, but the vanilla and Max U-Net fail to do so due to their 2D convolutional nature.





In addition to accuracy, inference speed is critical to enable model scalability and to check a large number of cycles rapidly. Table~\ref{tbl:run-times} compares the inference times of an industrial flow, PowerNet, MAVIREC, and baseline ML models. The runtimes are reported on D1 with 3.2 million instances and includes the inference and feature extraction time. Accounting for the feature extraction time allows a fair comparison against the industrial flow. The total runtime is 18 minutes, which gives a \textbf{10$\mathbf{\times}$ speedup} over the industrial flow rail analysis.  Most of the 18 minutes is spent in extracting the required features and the distribution of runtimes for each feature is listed in Fig.~\ref{fig:inference-flow}, and the inference alone is less than {\bf 3s}. MAVIREC is 2$\mathbf{\times}$ faster than max-U-Net and \textbf{100$\mathbf{\times}$ faster} than PowerNet in pure inference. The vanilla U-Net is faster than MAVIREC but is less efficient in predicting hotspots (Fig.~\ref{fig:pr}). 


\begin{table}
\centering
\caption{Runtime comparison of industrial flow and models.}
\label{tbl:run-times}
\resizebox{\linewidth}{!}{%
\begin{tabular}{|c|c|c|c|c|c|} 
\hline
 \textbf{Task}  & \begin{tabular}[c]{@{}c@{}}\textbf{Industrial}\\\textbf{~flow} \end{tabular} & \textbf{PowerNet}  & \begin{tabular}[c]{@{}c@{}}\textbf{Max }\\\textbf{U-Net} \end{tabular} & \begin{tabular}[c]{@{}c@{}}\textbf{Vanilla }\\\textbf{U-Net} \end{tabular} & \textbf{MAVIREC}  \\ 
\hline
\begin{tabular}[c]{@{}c@{}}\textbf{Feature }\\\textbf{extraction} \end{tabular} & \multirow{2}{*}{3 hours} & \multicolumn{4}{c|}{17 mins} \\ 
\cline{1-1}\cline{3-6}
\textbf{ML inference}  &  & 5 mins & 7.2s & 1s & 3s \\
\hline
\end{tabular}
}
\vspace{-0.2em}
\end{table}

\begin{figure}[h]
\centering
\includegraphics[width=8.5cm]{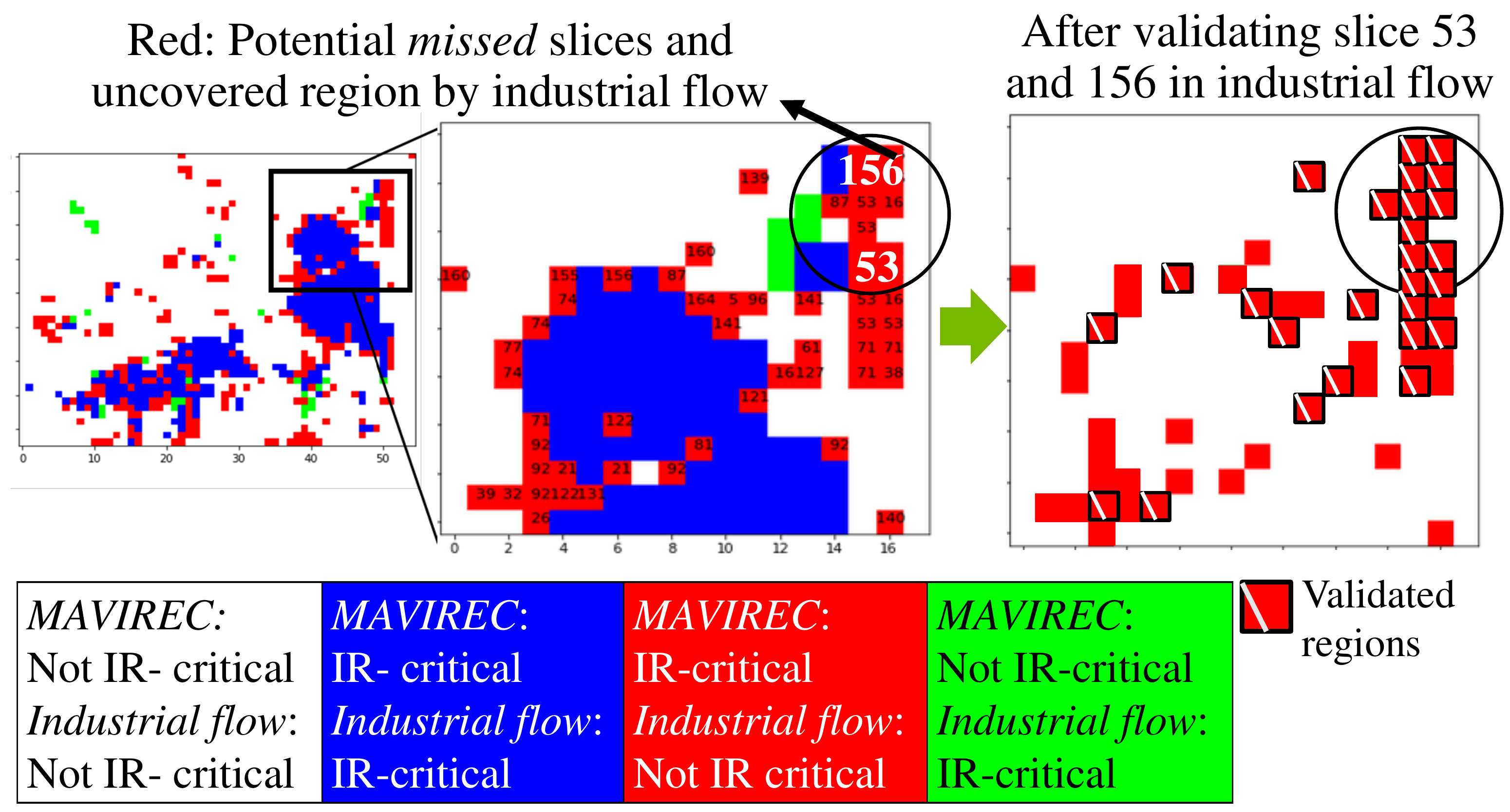}
\caption{Regional coverage comparison of MAVIREC's 168 candidate slices against 3 candidate slices from industrial flow. Validation of the largest industrial flow-uncovered IR-critical regions on design D2 as a representative example. }
\label{fig:vector-profiling-result}
\end{figure}

\subsection{MAVIREC for vector profiling}
\label{sec:results-vp}
\noindent
{\bf Quality of MAVIREC-recommended vectors}
The input vectors to our profiling algorithm (Section~\ref{sec:vector-profiling}) have about 100,000 clock cycles $\approx$ 5000 slices. For $N_a = 200$, $N_r =5$, $N_o = 3$, and $w \times l = 15\mu$m$\times15\mu$m ($6\times6$ tiles), the algorithm generates $N_c \approx 100$ candidate slices per vector. 
Fig.~\ref{fig:vector-profiling-result} shows the coverage comparison between the industrial flow-generated top-3 candidate slices and 168 MAVIREC-recommended slices on a representative design D2. Fig.~\ref{fig:vector-profiling-result} (leftmost) compares the regional coverage.
The middle picture shows the part of the design with the largest industrial flow-missed IR-critical spots (set of red regions). The numbers on the red regions indicate slice IDs (a numeric ID number for identifying each unique slice of the testbench) that resulted in those regions being reported as IR critical by MAVIREC. These are a set of \textit{missed} slices and we validate slice IDs 53 and 156 using an industrial flow. The rightmost map shows black-outlined and white-dashed red  regions which were validated to be IR-critical and the rest of the red regions are not validated. In the four designs we consider, each has $\approx$ 30 unique missed slices. It would be near-impossible to validate each missed slice by running industrial flow (ground truth) IR drop analysis, given storage (30GB per slice), time (3 hours per slice), and license limitations. Therefore, we limit our validation to missed slices with the largest uncovered region cluster.



Table~\ref{tbl:vector-profiling} lists the number of candidates, $N_c$, generated for each design for the test T1. For each of the designs we consider and for a region size of $6\times6$ tiles, we have 70--170 candidate slices generated while a industrial flow generated three slices. MAVIREC provides a large coverage by reporting an average of $\sim 5\%$ of the regions as potentially uncovered by the industrial flow and has less than 1.7\% false negatives.
\setlength{\tabcolsep}{3pt} 
\renewcommand{\arraystretch}{0.9} 

\begin{table}
\centering
\caption{MAVIREC vs. industrial flow recommended slices.}
\label{tbl:vector-profiling}
\resizebox{\linewidth}{!}{%
\begin{tabular}{||c||r|r|r|r|r|r|r|r||} 
\hhline{|t:=:t:==:t:==:t:====:t|}
\multirow{2}{*}{\begin{tabular}[c]{@{}c@{}} \textbf{Design,}\\\textbf{Test} \end{tabular}} & \multicolumn{2}{c||}{\begin{tabular}[c]{@{}c@{}}\textbf{Industrial flow }\\\textbf{profiling} \end{tabular}} & \multicolumn{2}{c||}{\begin{tabular}[c]{@{}c@{}}\textbf{MAVIREC }\\\textbf{profiling} \end{tabular}} & \multicolumn{4}{c||}{\textbf{Comparison} } \\ 
\cline{2-9}
 & \multicolumn{1}{c|}{\textbf{$N_c$} } & \multicolumn{1}{c|}{\begin{tabular}[c]{@{}c@{}}\textbf{\#IR-}\\\textbf{critical}\\\textbf{regions} \end{tabular}} & \multicolumn{1}{c|}{\textbf{$N_c$} } & \multicolumn{1}{c|}{\begin{tabular}[c]{@{}c@{}}\textbf{\#IR-}\\\textbf{critical }\\\textbf{regions} \end{tabular}} & \multicolumn{1}{c|}{\begin{tabular}[c]{@{}c@{}}\textbf{\#Regions}\\\textbf{reported}\\\textbf{IR-}\\\textbf{critical}\\\textbf{by both} \end{tabular}} & \multicolumn{1}{c|}{\begin{tabular}[c]{@{}c@{}}\textbf{\%Regions}\\\textbf{uncovered}\\\textbf{by}\\\textbf{MAVIREC} \end{tabular}} & \multicolumn{1}{c|}{\begin{tabular}[c]{@{}c@{}}\textbf{\%Regions}\\\textbf{uncovered}\\\textbf{by}\\\textbf{industrial}\\\textbf{flow} \end{tabular}} & \multicolumn{1}{c||}{\begin{tabular}[c]{@{}c@{}}\textbf{\#Unique }\\\textbf{slice IDs }\\\textbf{of missed}\\\textbf{regions} \end{tabular}} \\ 
\hhline{|:=:|--------||}
\textbf{D1, T1}  & 3 & 1163 & 133 & 1531 & 1093 & 1.7 & 10.8 & 30 \\ 
\hline
\textbf{D2, T1}  & 3 & 422 & 168 & 550 & 383 & 1.6 & 6.9 & 36 \\ 
\hline
\textbf{D3, T1}  & 3 & 369 & 166 & 553 & 329 & 1.7 & 9.4 & 26 \\ 
\hline
\textbf{D4, T1}  & 3 & 682 & 73 & 786 & 672 & 0.4 & 4.5 & 11 \\
\hhline{|b:=:b:========:b|}
\end{tabular}
}
\vspace{-0.7em}
\end{table}



\noindent
{\bf MAVIREC vector profiling runtimes} MAVIREC is able to provide high-quality recommendations for a 100K-cycle vector in 30 minutes while the industrial flow takes up to 2 hours.

\section{Conclusion}
\noindent
We propose MAVIREC: a fast and accurate vector profiling system that  recommends  a set of  worst-case  IR drop switching patterns using an ML-based IR drop estimation.  MAVIREC can profile hundred thousand-cycle vectors in under 30 minutes (4$\times$ speedup vs. industrial flows) on industrial designs and captures regions that were missed by industrial flows. MAVIREC's inference for IR drop estimation is 10$\times$  faster than industrial flows at the cost of a mere 4mV error. While this work focused on vectored dynamic IR drop analysis, the ML inference engine can be adopted for both vectorless dynamic and static IR drop analysis.

\bibliographystyle{IEEEtran}
\bibliography{references}
\end{document}